\documentclass[aps,prd,twocolumn,showpacs,eqsecnum,nofootinbib]{revtex4}

\usepackage{epsfig}
\usepackage{graphicx}
\usepackage{dcolumn}
\usepackage{amsmath}
\usepackage{enumerate}

\def\gtap{\mathrel{ \rlap{\raise 0.511ex \hbox{$>$}}{\lower 0.511ex
   \hbox{$\sim$}}}} 
\def\ltap{\mathrel{ \rlap{\raise 0.511ex
    \hbox{$<$}}{\lower 0.511ex \hbox{$\sim$}}}}

\begin{document}

\title{
\vskip-6pt \hfill {\rm\normalsize UCLA/04/TEP/5} \\
\vskip-12pt~\\
Low reheating temperature and the visible sterile neutrino}

\author{
\mbox{Graciela Gelmini$^{1}$},
\mbox{Sergio Palomares-Ruiz$^{1, 2}$} and
\mbox{Silvia Pascoli$^{1}$}}

\affiliation{
\mbox{$^1$  Department of Physics and Astronomy, UCLA,
  Los Angeles, CA 90095, USA}
\mbox{$^2$ Department of Physics and Astronomy, Vanderbilt University,
  Nashville, TN 37235, USA }
\\
{\tt gelmini@physics.ucla.edu},
{\tt sergiopr@physics.ucla.edu},
{\tt pascoli@physics.ucla.edu}}

\vspace{6mm}
\renewcommand{\thefootnote}{\arabic{footnote}}
\setcounter{footnote}{0}
\setcounter{section}{1}
\setcounter{equation}{0}
\renewcommand{\theequation}{\arabic{equation}}

\begin{abstract}
We present here a scenario, based on a low reheating temperature 
$ T_R \ll 100$~MeV at the end of
(the last episode of) inflation, in which the coupling 
of sterile neutrinos to active neutrinos can be as large as 
experimental bounds permit (thus making this neutrino ``visible'' 
in future experiments). In previous models 
this coupling was forced to be very small to prevent a cosmological
 overabundance of sterile neutrinos.
Here the abundance depends on how low the reheating temperature is.
For example, the sterile neutrino required by the LSND result does 
not have any cosmological problem within our scenario.
\end{abstract}

\pacs{14.60.St, 98.80.Cq}

\maketitle

In inflationary models, the beginning of the radiation dominated era of
the universe results from the decay of coherent oscillations of a
scalar field and the subsequent thermalization of the decay products
into a thermal bath with the so called ``reheating temperature" $T_R$.
This temperature may have been as low as 0.7 MeV~\cite{kohri} (a very
recent analysis strengthens this bound to $\sim$ 4
MeV~\cite{hannestad}). It is well known that a low reheating 
temperature inhibits the production of particles which are
non-relativistic or decoupled  at $T \ltap T_R$~\cite{giudice1}. The
final number density of active neutrinos starts departing from the
standard number for $T_R \ltap 8$~MeV but stays within 10\% of it for
$T_R \gtap 5$~MeV. For $T_R = 1$~MeV the number of tau- and muon-
neutrinos is about 2.7\% of the standard number. This would have
allowed one of the active neutrinos to be a warm dark matter (WDM)
candidate~\cite{giudice2}. Experimental bounds force now all three
mostly-active neutrino masses to be in the range of hot dark matter
(HDM), but we can use the same idea on sterile neutrinos. 

Sterile neutrinos without extra-standard model interactions are
produced in the early universe through their mixing  with active
neutrinos~\cite{barbieri}. Dodelson and Widrow~\cite{dodelson} (see
also Ref.~\cite{dolgov}) provided the first analytical calculation of
the production of  sterile neutrinos in the early universe, under the
assumption (which we maintain here) of a negligible primordial lepton
asymmetry. Fig.~2 of  Ref.~\cite{abazajian} shows that mostly-sterile
neutrinos produced in this manner have an acceptable abundance only  
if their mixing  with active neutrinos is very small, for example
$\sin^2 2 \theta < 10^{-7}$ for masses $m_s \gtap 1$~keV. In the
presence of a large lepton asymmetry, sterile neutrinos are produced
resonantly~\cite{resonant, abazajian} with a non-thermal spectrum
which favors low energies (``cool" dark matter candidate). 

The main idea of this letter is that the primordial abundance of
sterile neutrinos does not necessarily impose their mixing to active
neutrinos to be  as small as usually believed. We can, thus,  consider
sterile neutrinos of any mass and coupling, as long as other
experimental and cosmological bounds are satisfied. These neutrinos
could, therefore, be revealed in future experiments. Only for
simplicity we do not deal here with neutrinos heavier than 1~MeV or
with a large chemical potential, or with reheating temperatures  lower
than 5~MeV, but our idea clearly applies to all of these cases too
\cite{longpaper}. 

In Ref.~\cite{dodelson} it is shown that most of the sterile neutrinos
are produced at a temperature $T_{max} \simeq 133~(m_s / {\rm
  keV})^{1/3}$~MeV. Thus, if $T_R < T_{max}$ the production of 
sterile neutrinos is suppressed. We follow the calculations of
Ref.~\cite{dodelson}, but consider that the production of sterile
neutrinos, through the conversion of active neutrinos produced in
collisions, starts when the temperature of the universe is $T_R <
T_{max}$. 

In the calculation the active neutrinos are assumed to have the usual
thermal equilibrium distribution $f_A = (\exp{E/T} + 1)^{-1}$, thus,
following Ref.~\cite{giudice2}, we restrict ourselves to reheating
temperatures $T_R \geq 5 $~MeV. We also restrict ourselves to the case
of sterile neutrinos with mass $m_s < 1$~ MeV, so that we do not need
to consider their decays into electron pairs. These sterile neutrinos
are, therefore, relativistic at production. 

In the approximation of two-neutrino mixing, $\sin \theta$ is the
amplitude of the heavy mass eigenstate in the composition of the
active neutrino flavor eigenstate $\nu_{\alpha}$,
$\alpha=e,\mu,\tau$. 

For some  range of masses which depend on $T_R$, one can neglect all
matter effects, so the oscillations are as in the vacuum. For example,
for $T_{R} = 5$ MeV, the  specific value of the reheating temperature
we use in this letter, this happens for $m_s \gtap 0.2$ eV (0.1 eV)
for $\nu_e \leftrightarrow \nu_s$ ($\nu_{\mu,\tau} \leftrightarrow
\nu_s$). In this case, the $\nu_s$ distribution function turns out to
be   
\begin{equation}
f_s(E,T)\simeq 3.2~d_{\alpha}\left(\frac{T_{R}}{5~{\rm MeV}}\right)^3 
\sin^22\theta
\left(\frac{E}{T}\right) f_\alpha(E,T)
\label{distribution}
\end{equation}
where $d_{\alpha}= 1.13$ for $\nu_{\alpha}= \nu_e$ and $d_{\alpha}=
0.79$ for $\nu_{\alpha}= \nu_{\mu,\tau} $ \cite{fermi}. This
distribution results in a number fraction of sterile over active
neutrinos plus antineutrinos 
\begin{equation}
{\rm f} \equiv \frac{n_{\nu_s}}{n_{\nu_\alpha}} \simeq 10~d_{\alpha}~
\sin^22\theta \left(\frac{T_{R}}{5~{\rm MeV}}\right)^3~.
\end{equation}
Notice that the  number density of sterile neutrinos depends only on
the active-sterile mixing angle and the reheating temperature. A low
reheating temperature insures a small sterile number density, even for
very large active-sterile mixing angles, as large as other
experimental bounds permit. This makes sterile neutrinos in our
scenario potentially detectable in future experiments. The
$\nu_s$--number density is independent of the mass of the sterile
neutrinos, contrary to the result of Ref.~\cite{dodelson}. Thus, the
mass density of non-relativistic sterile neutrinos, $\Omega_sh^2 =
(m_s \; n_{\nu_s}/\rho_c) h^2$ depends linearly on the mass and on
$\sin^22\theta$, 
\begin{equation}
\Omega_sh^2 \simeq 0.1~d_{\alpha}~
 \left(\frac{\sin^22\theta}{10^{-3}}\right)            
 \left(\frac{m_s}{1~\rm{keV}}\right) 
\left(\frac{T_R}{5~\rm{MeV}}\right)^3~.
\end{equation}

\begin{figure}
\centerline{\epsfxsize=3.7in \epsfbox{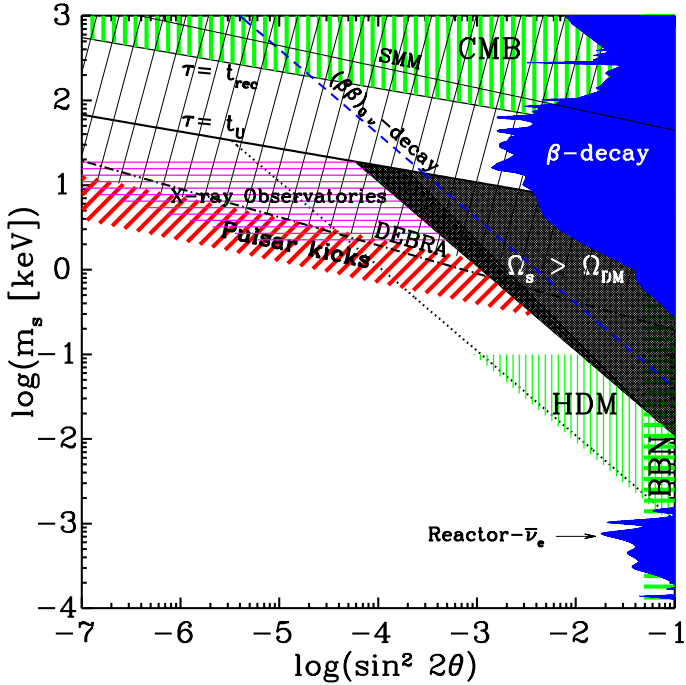}}
\caption{\label{nue} 
Bounds and sensitivity regions for $\nu_e \leftrightarrow \nu_s$
oscillations. See text.}   
\end{figure} 

\begin{figure}
\centerline{\epsfxsize=3.7in \epsfbox{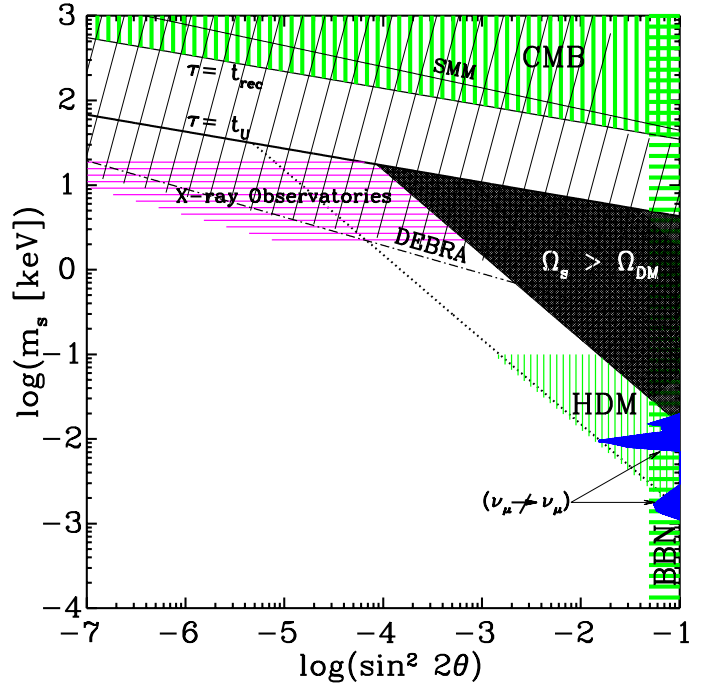}}
\caption{\label{numu} Same as Fig.~1 for $\nu_{\mu, \tau}
  \leftrightarrow \nu_s$. For $\nu_\tau  \leftrightarrow \nu_s$ the
 darkest gray-blue excluded region does not apply. See text.}    
\end{figure} 

The condition $\Omega_sh^2 \leq \Omega_{DM} h^2 = 0.135$~\cite{WMAP}
rejects the triangular dark gray region of masses and mixings shown in 
 Figs.~1  and 2. The values of masses and mixings for which sterile
 neutrinos constitute 10\% of the dark matter are also shown with a
 dotted line. 

Figs.~1 and 2  show  bounds for $\nu_{\alpha}= \nu_e$ and
$\nu_{\alpha}= \nu_{\mu,\tau}$, respectively, and for $T_R =5$~MeV. 
They show that $\nu_s$ in our scenario are viable HDM candidates,
while  neutrinos with $ m_{s} > 1$~keV are disfavoured, if not
rejected, as WDM or CDM, by bounds coming from supernovae cooling and
astrophysical bounds due to radiative decays, explained below.

Through $\nu_\alpha \leftrightarrow \nu_s$ oscillations, sterile
neutrinos can be produced in supernova (SN) cores and escape, carrying
away a large amount of the released energy. The observations of
$\nu_e,\bar{\nu}_e$ from SN1987A constrain the energy loss in $\nu_s$
and yield a bound on the mixing angle. For $m_{s} \ltap 45 ~{\rm keV}$,
$\nu_\alpha \leftrightarrow \nu_s$ oscillations are matter suppressed
and the forbidden range is~\cite{kainulainen}
\begin{equation}
\label{matter}
0.22  ~{\rm keV} \ltap |m_{s} (\sin^{2}{2 \theta})^{1/4}|
 \ltap 17~{\rm keV}~.
\end{equation}
For $m_{s} \gtap 45 ~{\rm keV}$, the matter effects are negligible and
neutrinos oscillate as in vacuum. In this case, the forbidden range is
\cite{kainulainen} 
\begin{equation}
\label{vacuum}
7 \times 10^{-10} \ltap \sin^{2}{2 \theta} \ltap 2 \times 10^{-2}
\end{equation}
These bounds exclude the diagonally hatched region with thin lines in
Figs. 1 and 2. The effective matter potential in the SN core for
$\nu_e \rightarrow \nu_s$ conversions, $V_m$, might be driven to its
zero equilibrium fixed point, $V_m \simeq 0$, during the
explosion~\cite{abazajian,FKMP}. In this case the $\nu_\alpha
\rightarrow \nu_s$ conversion happens as in vacuum. However, using the
SN parameters of Ref.~\cite{kainulainen}, this does not change the
bound in Eqs.~\ref{matter} and \ref{vacuum}. The SN1987A bound on
neutrino radiative decays~\cite{SMM}, excludes the region above the
line labeled  SMM (Solar Maximum Mission satellite) in both figures.

If the sterile neutrinos produced in non-resonant $\nu_e \rightarrow
\nu_s$ conversion, in fact, carry away a sizable fraction of the
energy emitted in a SN explosion, asymmetric emission of $\nu_s$ due
to the presence of a strong magnetic field, could explain the very
large velocities of pulsars~\cite{FKMP} (see diagonally hatched region
with thick lines in Fig. 1). 

Having restricted ourselves to $m_s < 1$~MeV, the dominant decay mode
of the mostly-sterile $\nu$ is into three neutrinos. Assuming
neutrinos are Majorana particles, a lifetime $\tau$ equal to the
lifetime of the universe $t_U =4.32\times 10^{17}$~s is indicated in
the figures with the full thick line (for Dirac neutrinos, the
lifetime is larger by a factor of $\sim 2$). Equivalent lines
corresponding to shorter or longer lifetimes can be easily obtained
knowing that the lifetime is proportional to $(\sin^2 2\theta ~
m_{s}^5)^{-1}$. The decay mode into a neutrino and a photon happens
with a branching ratio $ 0.8 \times 10^{-2}$ (this decay is not GIM
suppressed, contrary to the decay of an active neutrino into another
active neutrino and a photon~\cite{wolfenstein}). The diffuse
extragalactic background radiation (DEBRA) imposes a bound~\cite{debra}
that, in the relevant range of masses, can be well approximated by
\begin{equation}
I_\gamma \ltap \left(E/0.05~{\rm MeV}\right)^{-1}
\,\left(\rm{cm}^2~\rm{sr}~\rm{s}\right)^{-1} 
\label{DEBRA}
\end{equation}
\noindent
where $I_\gamma$ is the differential photon flux. 

For decays with $\tau >t_{rec}$ ($t_{rec}$ is the time of
recombination in the early universe; the line $\tau =t_{rec}$ is also
shown in the figures) the bound obtained for unclustered neutrinos
would reject the region above the dot-dashed line labeled DEBRA in the
figures. In particular for $\tau > t_U$, for unclustered sterile
neutrinos, the bound would be
\begin{equation}
\left(\frac{m_s}{1 ~\rm{keV}}\right) \ltap 0.10~d_{\alpha}^{-1/6}~
\left(\frac{5~{\rm MeV}}{T_{R}}\right)^{1/2}
\frac{1}{(\sin^22\theta)^{1/3}}~.
\end{equation}
However, this bound affects  neutrinos with mass $m_{s} \gtap 100$~eV
which are gravitationally clustered. Therefore the bound is not
entirely  correct. But the actual estimate of  how much of the diffuse 
photon background would be due to neutrinos which decay after
structures in the universe form, is missing in the literature.

Abazajian et al.~\cite{abazajian2} proposed to observe clusters of
galaxies with the Chandra and XMM-Newton observatories, in their high
sensitivity range for X-ray photon detection of 1--10~keV. They
proposed to reach a detection energy flux of $10^{-13}$~erg/(cm$^2$ s)
with the Chandra observatory, which would allow to observe a
monochromatic signal from the Virgo cluster, if 
\begin{equation}
\left(\frac{m_s}{2 ~\rm{keV}}\right)
 \gtap 2.1~d_{\alpha}^{-1/6}~ \left(\frac{5~{\rm MeV}}{T_{\rm
 R}}\right)^{1/2} \left(\frac{10^{-6}}{\sin^22\theta}\right)^{1/3},
\end{equation}
for $ m_s =$~ 2~--~20~keV (horizontally hatched region with thin lines
in the figures). Here the density fraction of sterile neutrinos within
clusters is assumed to  coincide with the cosmological energy fraction
($\Omega_s /\Omega_{DM}$).   

The lack of distortions in the CMB spectrum due to neutrino radiative
decays, excludes all the vertically hatched region with thick lines in
the figures~\cite{COBE}. 

Structure formation arguments impose sterile neutrinos which
constitute the whole of the dark matter (WDM) to have $m_{_s} >
2.9$~keV~\cite{hansen}. Note we use 2.9~keV instead of the 2.6~keV in
Ref.~\cite{hansen} because of our choice of cosmological
parameters. Besides, our sterile neutrinos are  hotter than those of
Ref.~\cite{hansen}, so the lower bound in our scenario should be even
somewhat larger than 2.9~keV. 

In the mass range in which sterile neutrinos can be part of the HDM,
we can apply the bounds on the sum of the contributions of active and 
sterile neutrinos to the HDM density. The 3$\sigma$ bound on the sum
of the neutrino masses (see Fig.~2 of Ref.~\cite{barger}) is 
\begin{equation}
\label{HDM}
\sum m_{i} + {\rm f} ~m_{s} \leq 1.1~{\rm eV}~,  \hspace{5mm}
\rm{i}=1,2,3 
\end{equation}
where $m_i$ are the light neutrino masses. Combining Eq.~\ref{HDM}
with an estimate of the light neutrino masses we obtain an upper limit
on $m_s$. If the  neutrino mass spectrum is normal hierarchical,
oscillation data impose the sum of the active neutrino masses to be
about 0.05~eV. This provides the most conservative bound on $m_s$:
\begin{equation}
m_s~\sin^2{2 \theta} \ltap 0.1~d_{\alpha}^{-1}~\left(T_R/5~{\rm
  MeV}\right)^{-3 }~\rm{eV}, 
\end{equation}
which we plot for $m_s \ltap 100$~eV (vertically hatched region with
thin lines in Figs.~1~and~2). 

The 3$\sigma$ upper bound imposed by big bang nucleosynthesis (BBN) on
 any extra contribution to the energy density, parametrized as extra
neutrino species, $\Delta N_\nu$, is $\Delta N_\nu\leq 0.73$ (see
 Fig.~7 of Ref.~\cite{steigman}), which translates into (horizontally
 hatched region with thick lines in Figs.~1 and 2.)
\begin{equation}
\sin^22\theta \ltap 5.6~d_{\alpha}^{-1}~10^{-2}~\left(T_R/5~{\rm
  MeV}\right)^{-3 }~.
\end{equation}
Let us turn now to experimental bounds. So far disappearance
experiments have reported negative
results~\cite{CHOOZPV,Bugey,CDHS}. Reactor-$\bar{\nu}_e$ experiments
CHOOZ~\cite{CHOOZPV} and Bugey \cite{Bugey} constrain the mixing angle
relevant in $\bar{\nu}_e \rightarrow \bar{\nu}_s$
conversion~\cite{CHOOZPV} to be $\sin^2{2 \theta} < 0.1$, for  $m_{s}
> 1.7$~eV. For smaller masses (down to $m_{s} \sim$ few $10^{-1}$ eV)
the bound is stronger by a factor of 2--5 \cite{Bugey} (darkest region
in Fig. 1). Accelerator-$\nu_\mu$ disappearance
experiments~\cite{CDHS} impose the mixing angle which controls
$\nu_\mu \rightarrow \nu_s$ oscillations to be  $\sin^2{2 \theta}\! <
\!0.02$, for $13.8~{\rm eV} \! <  \! m_{s} \! <  \! 17.9~{\rm
  eV}$. Less stringent bounds apply for other values of  $m_{s}$
(darkest region in Fig. 2). Future experiments looking for
$\bar{\nu}_e$~\cite{reactor} and $\nu_\mu$, $\bar{\nu}_\mu
$~\cite{MiniBooNE,MM} disappearance might explore part of the now
allowed parameter space, e.g., $\nu$-factories may reach $\sin^{2}{2
  \theta} \sim 10^{-3}$.  

Appearance experiments searching for $\nu_\alpha \rightarrow
\nu_\beta$ oscillations ($\alpha,\beta = e, \mu, \tau$), are
sensitive to the product of the mixing angles between $\nu_\alpha$,
$\nu_\beta$ and the mostly-sterile mass eigenstate. These experiments
have reported no positive signal, except for the LSND
experiment~\cite{LSND}, which found evidence of $\bar{\nu}_\mu
\rightarrow \bar{\nu}_e$ conversion. MiniBooNE~\cite{MiniBooNE} will
test this result. Let us notice that in our model, the ranges of
$m_s$ and $\sin^{2}{2 \theta}$ required to explain the LSND data (in
terms of neutrino oscillations), are cosmologically and
astrophysically allowed. The analysis of the data requires the mixing
of, at least, four neutrinos, and therefore cannot be used to set
bounds on the mixing angle $\sin^{2}{2 \theta}$ we have used to
parametrize 2--$\nu$ oscillations~\cite{longpaper}.  

In the mass range of interest, $\beta$--decay experiments searching
for kinks in the energy spectra of the emitted electron constrain the
mixing angle between $\nu_e$ and the mostly-sterile neutrino mass
eigenstate. Different nuclei have been used and negative results have
been found so far (for a complete review see Ref.~\cite{PDG}). The
limits are strongly mass dependent (darkest region in Fig.~1). 

If neutrinos are Majorana particles, the neutrinoless double beta
($(\beta \beta)_{0 \nu}$) --decay is allowed. The half-life time
depends on the effective Majorana mass $\langle m \rangle$ (see
e.g. Ref.~\cite{BiPet87}). The contribution of the mostly-sterile
neutrino is of the form $\langle m \rangle_s = m_s \, \sin^2{\theta}
\, e^{i \beta_s}$ where $\sin^2{2 \theta}$ is the mixing parameter in
$\nu_e \rightarrow \nu_s$ conversions and $\beta_s$ is a Majorana
CP--violating phase. At present, the most stringent bound on $|\langle
m \rangle|$ is $|\langle m \rangle| < 0.35-1.05
~\rm{eV}$~\cite{76Ge00}, which conservatively translates into (dashed
line in Fig.~1): 
\begin{equation}
 m_s  \ \sin^2{2 \theta} < 4 ~ {\rm eV} ~.
\end{equation}
Possible cancellations in $|\langle m \rangle|$ among contributions
due to different mass eigenstates would weaken this bound. Barring
this possibility, the present and future $(\beta \beta)_{0
  \nu}$--decay experiments (see, e.g., Ref. \cite{AviNoon2004}) will
probe part of the cosmologically relevant region, possibly up to $m_s
\; \sin^{2}{2 \theta} \sim 0.05 \; \rm{eV}$, in which $\nu_s$ could be
an important part of the dark matter or produce pulsar kicks.

The sterile neutrinos with $\sin^{2}{2 \theta} \approx$ 0.1 -- 0.01
would have the cross section required in Ref.~\cite{Barbot} to
separate atmospheric showers produced by these neutrinos from showers
generated by active neutrinos, in future experiments such as EUSO and
OWL~\cite{eusoowl}, by using the Earth as a filter. The required  flux
of ultra-high energy neutrinos  would be very large.

We presented here a scenario, based on a low reheating temperature at
the end of inflation, in which the coupling of sterile neutrinos to
active neutrinos can be as large as experimental bounds permit. For
example, the sterile neutrino required by the LSND result does not
have any cosmological or astrophysical problem. 

The experimental discovery of a sterile neutrino in the region of
$m_s-\sin^2 {2\theta}$ opened up in this paper, would require an
unusual cosmology, such as one with a low reheating temperature as
presented here. In this case, baryon asymmetry might be generated
through the Affleck-Dine mechanism~\cite {Affleck}, and the bulk of
the dark matter (if not made of sterile neutrinos) should consist of
other non-thermally produced particles. 

We thank A. Kusenko and D. Semikoz for discussions. This work was
supported in part by the DOE grant DE-FG03-91ER40662, Task C and NASA
grant NAG5-13399.

\end{document}